\documentstyle[11pt,newpasp,twoside, epsf]{article}
\def\edcomment#1{\iffalse\marginpar{\raggedright\sl#1\/}\else\relax\fi}
\reversemarginpar

\begin{document}
\title{Quest for HI Turbulence Statistics: New Techniques}
 \author{Alex Lazarian}
\affil{University of Wisconsin-Madison, USA; {\it lazarian@astro.wisc.edu}}
\author{Dmitry Pogosyan}
\affil{University of Alberta, Canada; {\it pogosyan@Phys.UAlberta.CA}}
\author{Alejandro Esquivel}
\affil{University of Wisconsin-Madison, USA; {\it esquivel@astro.wisc.edu}}

\begin{abstract}
HI data cubes are sources of unique information on interstellar turbulence. 
Doppler shifts due to supersonic motions contain information on 
turbulent velocity field which is otherwise difficult to obtain.
However, the problem of separation of velocity and density fluctuations 
within HI data cubes is far from being trivial. Analytical
description of the emissivity statistics of channel maps (velocity slices)
in Lazarian \& Pogosyan (2000) showed that the relative contribution of the
density and velocity fluctuations 
depends on the thickness of the velocity slice.
In particular, power-law assymptotics of the emissivity fluctuations change
when the dispersion of the velocity at the scale under study 
becomes of the order of the velocity slice thickness (integrated
width of the channel map).   
These results are the foundations of the Velocity-Channel Analysis (VCA) 
technique which allows to determine velocity and density statistics
using 21-cm data cubes. The VCA has been successfully tested using data
cubes obtained via compressible magnetohydrodynamic simulations and applied
to Galactic and Magellanic Clouds data. As a tool it has become much more
sophisticated recently when effects of absorption were accounted for.
The systematic studies of vast 21-cm data sets to 
correlate the variations in the turbulence statistics
with the astrophysical activity is on the agenda. This should
allow to determine the interstellar energy injection mechanisms. Going
beyond the VCA, we discuss other tools, namely, genus and anisotropy
analysis. The first characterises the topology of HI, while the
second provides magnetic field directions.
We show a few applications of these new tools to HI data and MHD 
simulations.     
\end{abstract}

\section{Introduction}

Atomic hydrogen is an important component of the interstellar media 
of spiral galaxies
(McKee \& Ostriker 1977) and much
efforts have been devoted to its studies (see this volume). From the point
of view of fluid mechanics, HI, as well as other components of interstellar
medium, is characterised by huge Reynolds numbers, $Re$, which is  
the ratio of the
eddy turnover time of a parcel of gas to the time required for viscous
forces to slow it appreciably. For $Re\gg 100$ we expect gas to be
turbulent and this is exactly what we observe in HI (for HI $Re\sim 10^8$).

Statistical description is a nearly indispensable strategy when
dealing with turbulence. The big advantage of statistical techniques
is that they extract underlying regularities of the flow and reject
incidental details. 
Attempts to study interstellar turbulence with statistical tools
date as far back as the 1950s
(see Horner 1951, Kampe de Feriet 1955, Munch 1958, 
Wilson et al. 1959) and various directions
of research achieved various degree of success (see reviews by
Kaplan \& Pickelner 1970, Dickman 1985, Lazarian 1992, Armstrong, Rickett
\& Spangler 1995). 
Studies of turbulence statistics of ionized media were successful
(see Spangler \& Gwinn 1990) and provided the information of
the statistics of plasma density at scales $10^{8}$-$10^{15}$~cm. 
This research profited
a lot from clear understanding of processes of scintillations and scattering
achieved by theorists (see Narayan \& Goodman 
1989). At the same time 
the intrinsic limitations of the scincillations technique
are due to the limited number of sampling directions, relevance only to
ionized gas at extremely small scales, 
and impossibility
of getting velocity (the most important!) statistics directly.

``Seeing through dust'' at 21 cm potentially provides an enormous leap
for turbulence studies. First of all, statistics of HI line carries
the velocity information. Secondly, the statistical samples are extremely
rich and not limited to discrete directions. Thirdly, HI emission
allows to study turbulence at  large scales which have direct
relation to the scales of star formation and energy injection.

Deficiencies in the theoretical description has been, to our mind, the
major impediment to studies of turbulence using 21-cm data cubes. 
In particular, Crovisier \& Dickey (1983) 
Green (1993) measured the spectrum of 21-cm
intensity fluctuations, but it was unclear what those spectra mean.

Sally Oey (this volume) identifies several known problems of studying
HI turbulence. They are ({\it a}) integration of fluctuations 
along lines of sight, ({\it b}) 
contribution of velocity and density to the emissivity statistics
and ({\it c}) effects of absorption. The problems related to ({\it a})
were dealt in Lazarian (1995), while the issues ({\it b}) and ({\it c})
happen to be more formidable and have been dealt with only recently (Lazarian
\& Pogosyan 2000, 2002). Earlier reviews dealing with HI studies
include Lazarian (1999a,b). In this volume one can read more
about turbulence in Cho, Lazarian \& Yan and Vazquez-Semadeni papers.

\section{Statistics of HI Data Cubes and Velocity Channel Analysis}

Doppler shifted spectral lines carry information about turbulent 
velocity, but until very recently there was no way of extracting this 
information. The problem is that both velocity and density 
fluctuations contribute to the fluctuations of the intensity at a 
given velocity. In other words: {\it We know the statistics of spectral
lines, i.e. the statistics in velocity space, while we want to determine the 
actual 3D statistics.}

{\bf Statistics in 3D}\\
Turbulence theory deals with the statistics of turbulence in real space that
can be described by 3D correlation and structure functions (see 
Monin \& Yaglom 1975). For instance, 
if density correlation function  (CF) is isotropic in $xyz$-space then
\begin{equation}
\xi(r)=\xi({\bf r}) = \langle \rho ({\bf x}) \rho ({\bf x}+{\bf r}) \rangle~~
~. 
\label{xifirst}
\end{equation}
The power spectrum provides an alternative description and is related to
the correlation function as
\begin{equation} 
P({\bf k})=\int d {\bf r} e^{i{\bf k}{\bf r}} \xi(\bf r)~~~,
\label{spectrum:gen}
\end{equation}
where integration is performed over the 3D space.
For power-law statistics,
the $N$ dimensional power spectrum  $P_N\sim k^{n}$
and correlation function $\xi_N\sim r^{-\gamma}$
have indexes related as $n=-N+\gamma$. In other words, 
\begin{equation}
({spectral~index})= (-{dimensions~of~space}+{CF~index})
\end{equation}

{\bf Statistics in Position Position Velocity (PPV) Space}\\
One does not observe the gas distribution in the real space galactic
coordinates ${\bf x}$. Rather, intensity of the emission in a given spectral
line is measured towards some direction on the sky and at a
given line-of-sight velocity $v$.
In the plane parallel approximation the direction on the sky can be identified
with the 2D spatial vector ${\bf X}$ perpendicular to the line of sight,
so that the coordinates of experimental PPV cubes are $({\bf X},v)$.
If we chose galactic coordinates so that $x,y$ specify the direction of 
observation and $z$ is the coordinate along the line of sight 
${\bf x}=({\bf X},z)$, then the relation between the 
real space and the PPV descriptions
is that of a map $({\bf X},z) \to ({\bf X},v)$.

The line-of-sight component of velocity $v$ at the position ${\bf x}$
is a sum of the 
regular gas flow (e.g., due to galactic rotation) $v_{gal}({\bf x})$,
the turbulent velocity $u({\bf x})$ and the residual component
due to thermal motions. The thermal velocity
$v_{thermal}=v-v_{gal}({\bf x})-u({\bf x})$ 
has Maxwellian distribution
\begin{equation}
\phi_v({\bf x}) {\mathrm d}v =\frac{1}{(2\pi \beta)^{1/2}}
\exp\left[-\frac{(v-v_{reg}({\bf x})-u({\bf x}))^2}
{2 \beta }\right] {\mathrm d} v ~~~,
\label{phi}
\end{equation}
where $\beta=\kappa_B T /m$, $m$ is the atomic mass and $T$ is the gas 
temperature.

The density of emitters in the PPV space 
$\rho_s({\bf X},v)$ can then be written as
\begin{equation}
\rho_s({\bf X},v){\mathrm d {\bf X} d} v = 
\left[ \int {\mathrm d} z\; \rho({\bf x}) \phi_v({\bf x}) \right]
{\mathrm d {\bf X} d} v 
\label{rhoz}
\end{equation} 
where $\rho({\bf x})$ is the density of gas
in the galactic coordinates.
This expression just counts the number of atoms
along the line-of-sight {\bf X}
which have z-component of velocity in the interval
$[v,v+dv]$ 
The limits of integration are defined by the extend of spatial
distribution of emitting gas. 

The statistics available in PPV are the correlation functions\\
  $\xi_s = \langle \rho_s({\bf X}_1, v_1)
\rho_s({\bf X}_2,v_2)\rangle$ and the power spectra of the emissivity fluctuations.
The latter may be 3D, 2D and 1D. The relation between those spectra and the
underlying velocity and density spectra were established in Lazarian \& Pogosyan (2000). In practical
terms, the most frequently used is the 2D spectrum of velocity slices. This
spectrum can be directly obtained with radiointerferometric observations (see
Lazarian 1995).  

{\bf Velocity-Channel Analysis (VCA)}\\
The VCA technique has been developed 
in Lazarian \& Pogosyan (2000, henceforth LP00)
for analyzing line 
profile formation in the presence of turbulence. We have predicted 
analytically that the relative contribution of the velocity and density 
fluctuations to the total fluctuations of intensity changes
in a regular fashion with the width of the velocity slice.
This is easy to understand
qualitatively, as it is clear that the line integration should decrease
the influence of velocity fluctuations. We have shown
 that if the 3-D density spectrum 
is $P_n \propto k^{n}$ (where $k$ is the wavenumber), 
two distinct regimes are present when: (a) $n>-3$, and (b) $n<-3$.

Our results are summarized in
Table~1. 
\begin{table*}
\caption{\label{t:lazarian+pogosyan}
A summary of analytical results derived in LP00.}
%\centering
\begin{tabular}{lcc}
\noalign{\smallskip} \hline \noalign{\smallskip} 
Slice & Shallow 3-D density & Steep 3-D density\\
thickness & $P_{n} \propto k^{n}$, $n>-3$&$P_{n} \propto k^{n}$, $n<-3$\\
%      & ($P_{2\rho}$ part dominates) &($P_{2v}$ part dominates)\\
\noalign{\smallskip} \hline \noalign{\smallskip}
2-D intensity spectrum for thin\tablenotemark{a}~~slice & 
$\propto k^{n+m/2}$    & $\propto 
k^{-3+m/2}$   \\
2-D intensity spectrum for thick\tablenotemark{b}~~slice & $\propto k^{n}$    
& $\propto k^{-3-m/2}$  \\
2-D intensity spectrum for very thick\tablenotemark{c}~~slice & $\propto k^{n}$ & $\propto k^{n}$  \\
\noalign{\smallskip} \hline \noalign{\smallskip}
\end{tabular}
\tablenotetext{a}{channel width $<$ velocity dispersion at the scale under
study}
\tablenotetext{b}{channel width $>$ velocity dispersion at the scale under
study}
\tablenotetext{c}{substantial part of the velocity profile is integrated over}
\end{table*}
It is easy to see that both
in cases (a) and (b) the power law index
{\it gradually steepens} with the increase of velocity slice
thickness. In the thickest velocity slices the velocity information
is averaged out and it is natural that we get the
density spectral index $n$. The velocity fluctuations dominate in 
thin slices, and if the 3-D velocity power spectrum is $P_v\sim k^{-3-m}$,
then the index $m$ can be found from thin slices (see Table~1). Note, that the 
notion of thin and thick slices depends on a turbulence scale under
study and the same slice can be thick for small scale turbulent fluctuations
and thin for large scale ones. The formal criterion for the slice to be
thick is that {\it the dispertion of turbulent velocities on the scale studied
should be less than the velocity slice thickness}.  Otherwise
the slice is {\it thin}.

Predictions of LP00 were tested in Lazarian et al. (2001) 
using numerical MHD simulations of compressible intersterllar gas.
Simulated data cubes allowed both density and velocity statistics
to be measured directly. Then these data cubes were
used to produce synthetic spectra which were analysed using the
VCA. As the result, the velocity
and density statistics were successfully recovered.
Thus one can confidently apply  the
technique to HI observations. Note, that the VCA 
is only sensitive to velocity\footnote{For Kolmogorov turbulence the
velocity gradient scales as $l^{-2/3}$.}  
and density gradients on the scale under 
study, so the regular Galactic rotation curve or large scale 
distribution of emitting gas does not interfere with the analysis. 

Further research in Lazarian \& Pogosyan (2002) incorporated the
effects of absorption into the VCA. Namely, we have shown that in the presence
of strong absorption 
a universal spectrum with index $-3$ is being
produced in very thick slices, provided that the density is steep. 
 The universality means that the emissivity spectral index does not
depend on the underlying statistics of velocity and density. For shallow
density, its spectral index determines the emissivity statistics.
To get velocity information thin slices must be used.

\section{Applications of the VCA: SMC and Galactic HI}

In the 
case of SMC the VCA reveals the Kolmogorov-type 
spectrum, $E(k)\equiv 4\pi k^2 P(k)\propto k^{-5/3}$ 
(where $E(k)$ is the energy contained 
in eddies with $k\sim$ (eddy size)$^{-1}$) for velocity fluctuations
and a 
slightly more shallow spectrum of density fluctuations
over the range of scales from 4 kpc to 
40 pc. 
\begin{figure} [h!t] 
{\centering \leavevmode 
\epsfxsize=2.5in\epsfbox{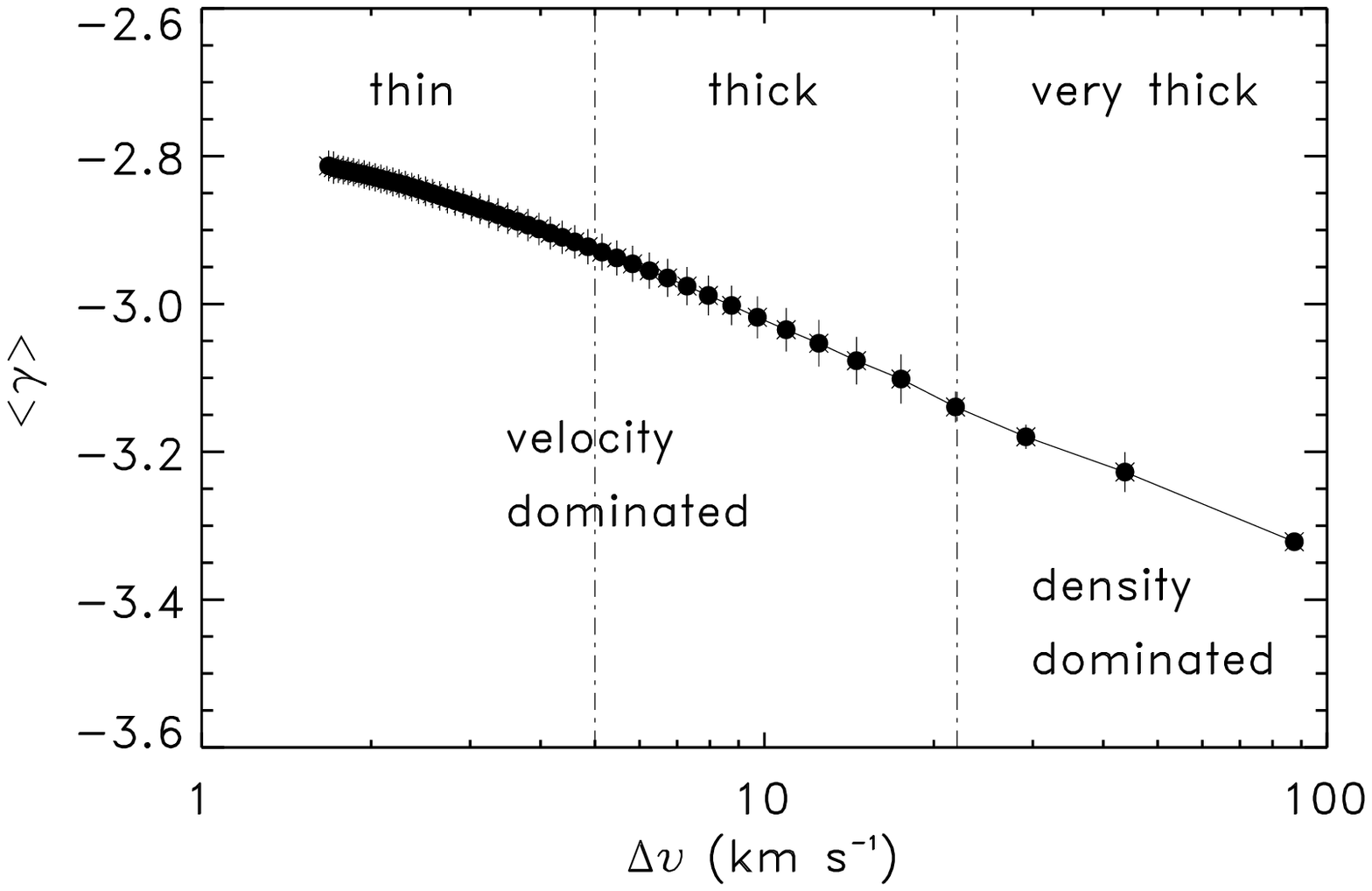} 
\hfil 
\epsfxsize=2.5in\epsfbox{ 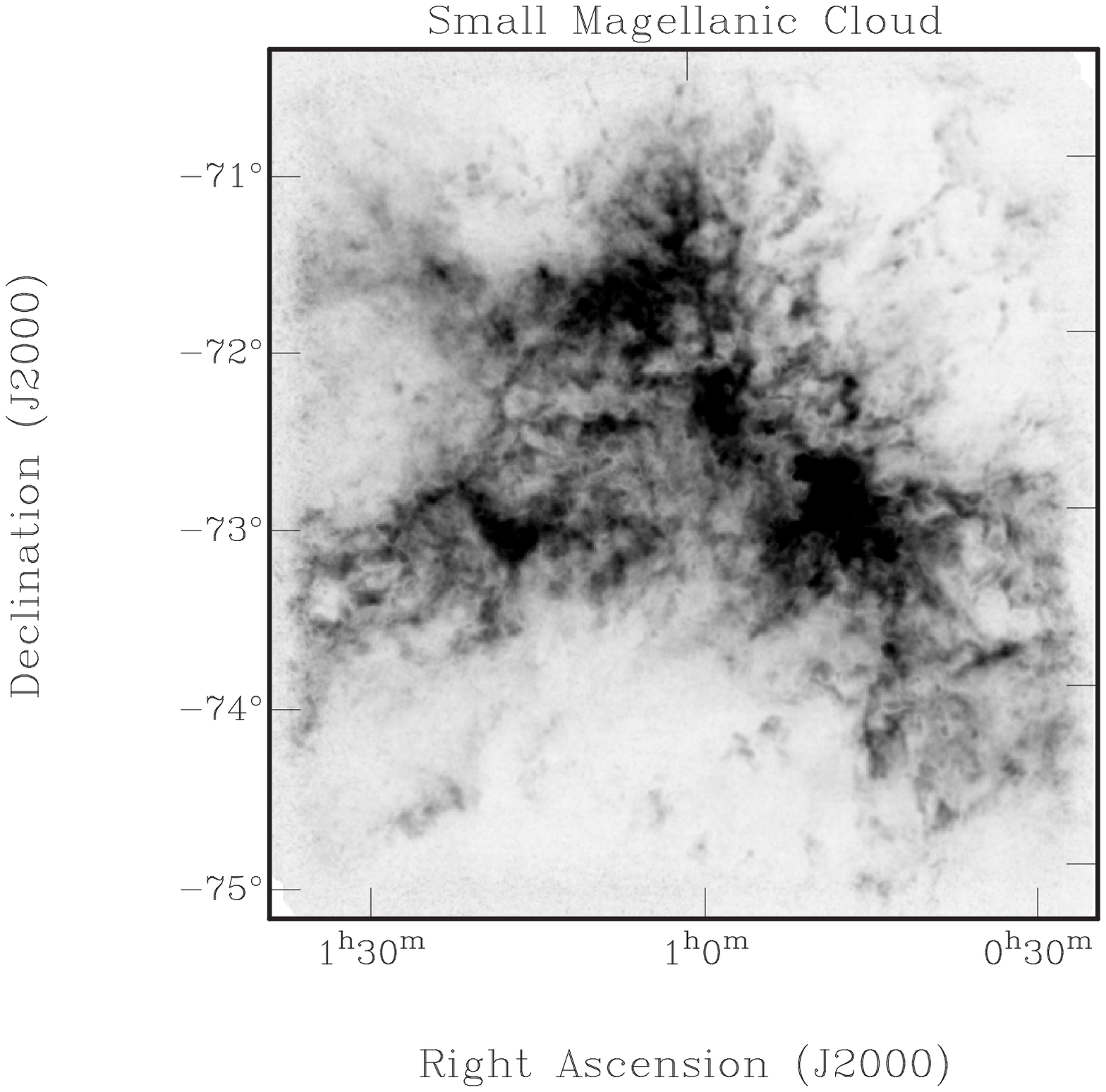} 
} 
\caption{{\it Left Panel}: Variations of two dimensional 21~cm 
spectral slope with the velocity slice thickness (from Stanimirovic \&
Lazarian 2001). The LP00 study predicts that the thick slice reflects
the density statistics, while the thin slice is influenced by the
velocity, which results in flattening of the slope.
{\it Right panel}: The 21~cm image of SMC that exhibits a lot
of density structure which, however, provides a subdominant
contribution to thin channel maps.}  
\end{figure}

A spectral index similar to the one for SMC has been inferred from the 21~cm
Galactic data in Green (1993). However, that index shows more variability,
which can stem from a number of causes. First of all, LP00 predicts,
and the analysis of Fig.~1 confirms,
that the emissivity index depends on whether or not
the velocity dispersion on the scale under study is larger or smaller
than the thickness of the channel map. In Green (1993) the physical
scales and therefore the associated velocity dispersion vary considerably
for close and distant slices of HI. Moreover, for the close slices
the geometry of parallel lines of sight adopted in LP00 does not
provide a good approximation and the convergence of the lines of sight
must be accounted for. 

The situation is even more complicated for the inner parts of the Galaxy,
where (a) two distinct regions at different distances from the observer
contribute to the emissivity for a given velocity and (b) effects of
the absorption are important. However, the analysis in Dickey et al. (2001) 
showed that some progress may be made even in those unfavorable
circumstances. Dickey et al. (2001) found the steepening 
 of the spectral index with the increase of the velocity slice thickness.
They also observed the spectral index for strongly absorbing direction
approached $-3$ in accordance with the conclusions in Lazarian \& Pogosyan
(2002).

21-cm absorption provides another way of probing turbulence on small
scales. The absorption depends on the density to temperature ratio
$\rho/T$, rather than to $\rho$ as in the case of emission. However,
 in terms of the VCA this change is not important and we still expect to
see emissivity index steepening as velocity slice thickness increases,
provided that velocity effects are present. In view of
this, results of Deshpande et al. (2001), who did not see such steepening,
can be interpreted as the evidence of the viscous suppression of
turbulence on the scales less than 1~pc. The fluctuations in this
case should be due to density and their shallow spectrum $\sim k^{-2.8}$ may
be related to the damped magnetic structures below the viscous
cutoff (see Cho, Lazarian, Yan, this volume).

Clearly, more VCA studies of Galactic HI are required.
As the VCA can be used for other emission lines (e.g. CO, H$_{\alpha}$), 
cross-correlation
of turbulence properties in HI, molecular and ionized gas is necessary to
obtain an insight into the dynamic coupling of different interstellar
phases.  

\section{Complementary Statistical Tools}

~~~~~~{\bf Genus Analysis}\\
Velocity and density power spectra do not provide the complete description
of turbulence. Intermittency of turbulence (its variations in time and space) 
and its topology in the presence of different phases are 
not described by the power spectrum.
 
``Genus analysis" is a good tool for studying the topology of
turbulence (see review by Lazarian 1999), receiving well-deserved
recognition for cosmological studies (Gott et al. 1989). Consider an
area on the sky with contours of projected density. The 2D genus,
$G(\nu)$, is the difference between the number of regions with a
projected density higher than $\nu$ and those with densities lower
than $\nu$. Fig.~4 shows the 2D genus as the function of $\nu$
for a Gaussian distribution of densities (completely symmetric curve),
for MHD isothermal
simulations with Mach number 2.5, and for HI in SMC (Fig. 4b).
\begin{figure} [h!t] 
{\centering \leavevmode 
\epsfxsize=2.5in\epsfbox{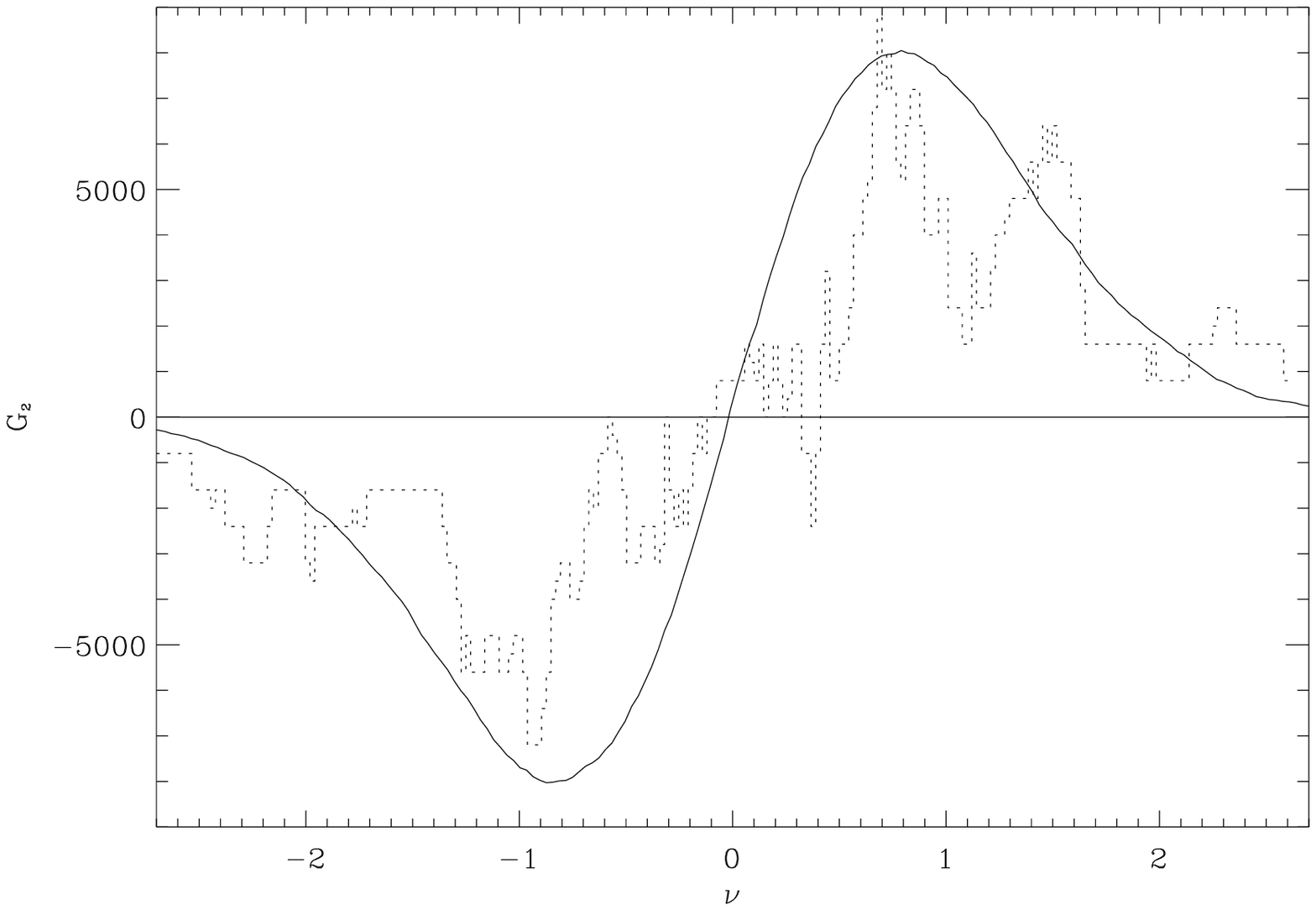} 
\hfil 
\epsfxsize=2.5in\epsfbox{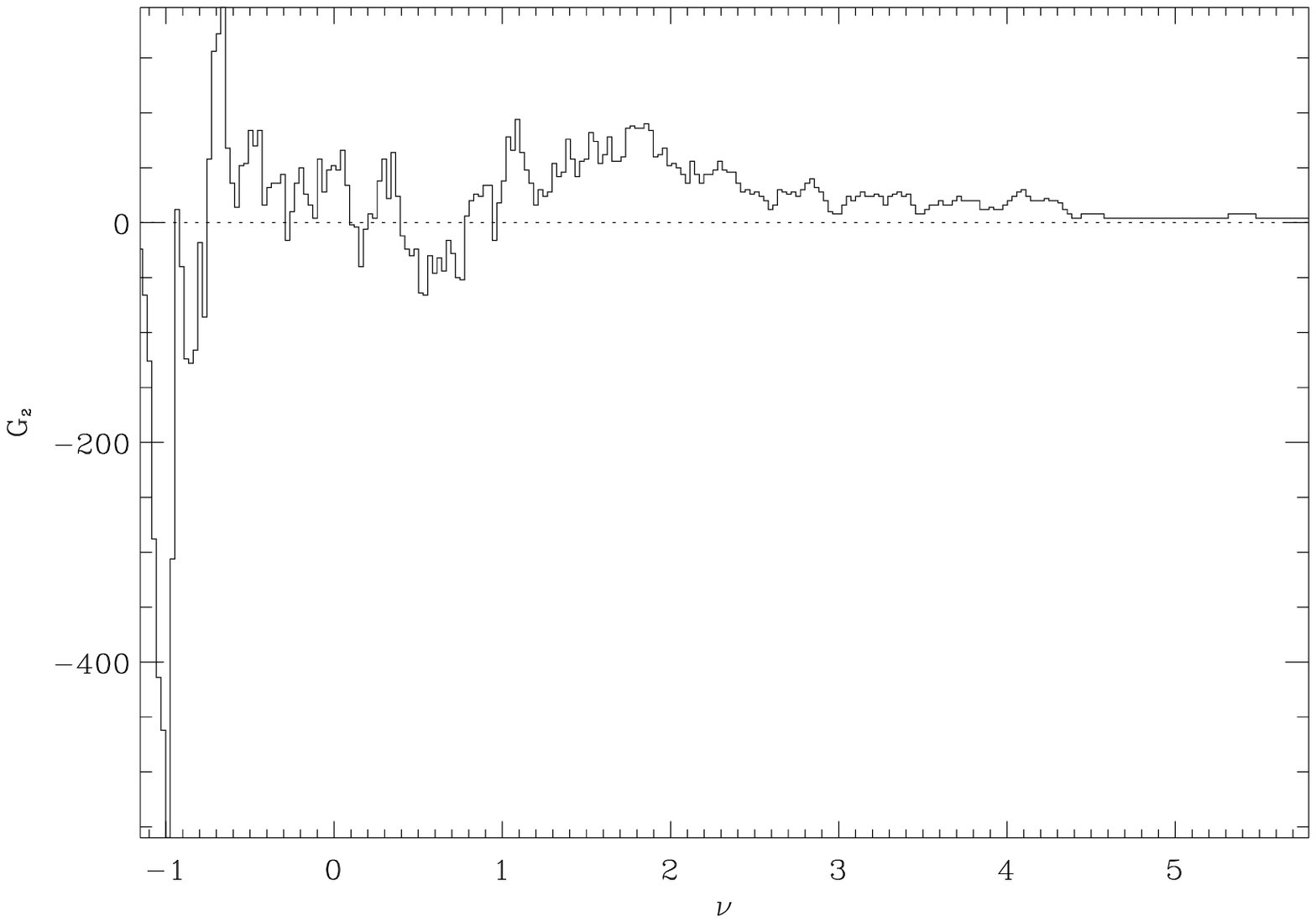} 
} 
\caption{{\it Left panel} (Fig4a) shows the 2D genus of the Gaussian 
distribution 
(smooth analytical curve) 
against the genus for the isothermal compressible MHD simulations with Mach 
number 2.5 (dotted curve). {\it Right panel} (Fig4b) shows the genus of 
HI distribution in SMC. 
The topology of the HI distribution is very different, although the 
spectra are similar (from Esquivel, Lazarian, Pogosyan
\& Cho, in preparation).} 
 \end{figure} 
 
Isothermal MHD simulations exhibit more or less symmetric density
distributions, but the SMC data reflect a prominent Swiss cheese
topology, which also can be suggested from the visual inspection of
the image (see Fig.~3). However, unlike visual inspection, the genus quantifies the
topology and allows us to compare numerical results with
observations. Note, that the MHD simulations in Fig~4a are not so different
from the SMC in terms of the power spectrum.

{\bf Anisotropy Analysis}\\
In isotropic turbulence, correlations depend only on the distance
between the points. Contours of equal correlation are 
circular in this case. Presence of magnetic field introduces anisotropy
and these contours become elongated with a symmetry axis given by
the magnetic field.  To study turbulence
anisotropy, we can measure contours of equal correlation corresponding
to the data within various
velocity bins. The results obtained with simulated data are shown in Fig.~3. 
\begin{figure} [h!t] 
{\centering \leavevmode 
\epsfxsize=2.5in\epsfbox{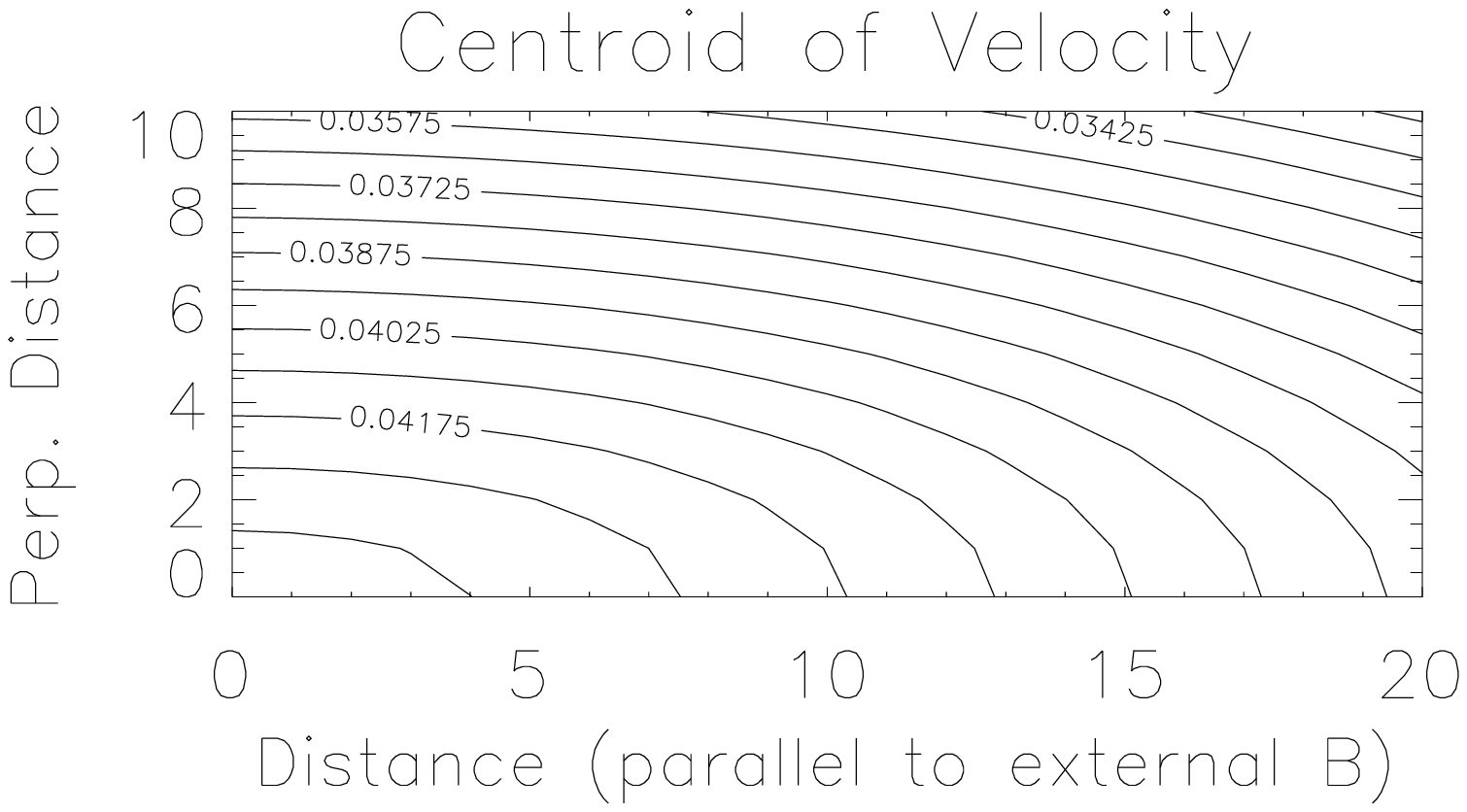} 
\hfil 
\epsfxsize=2.5in\epsfbox{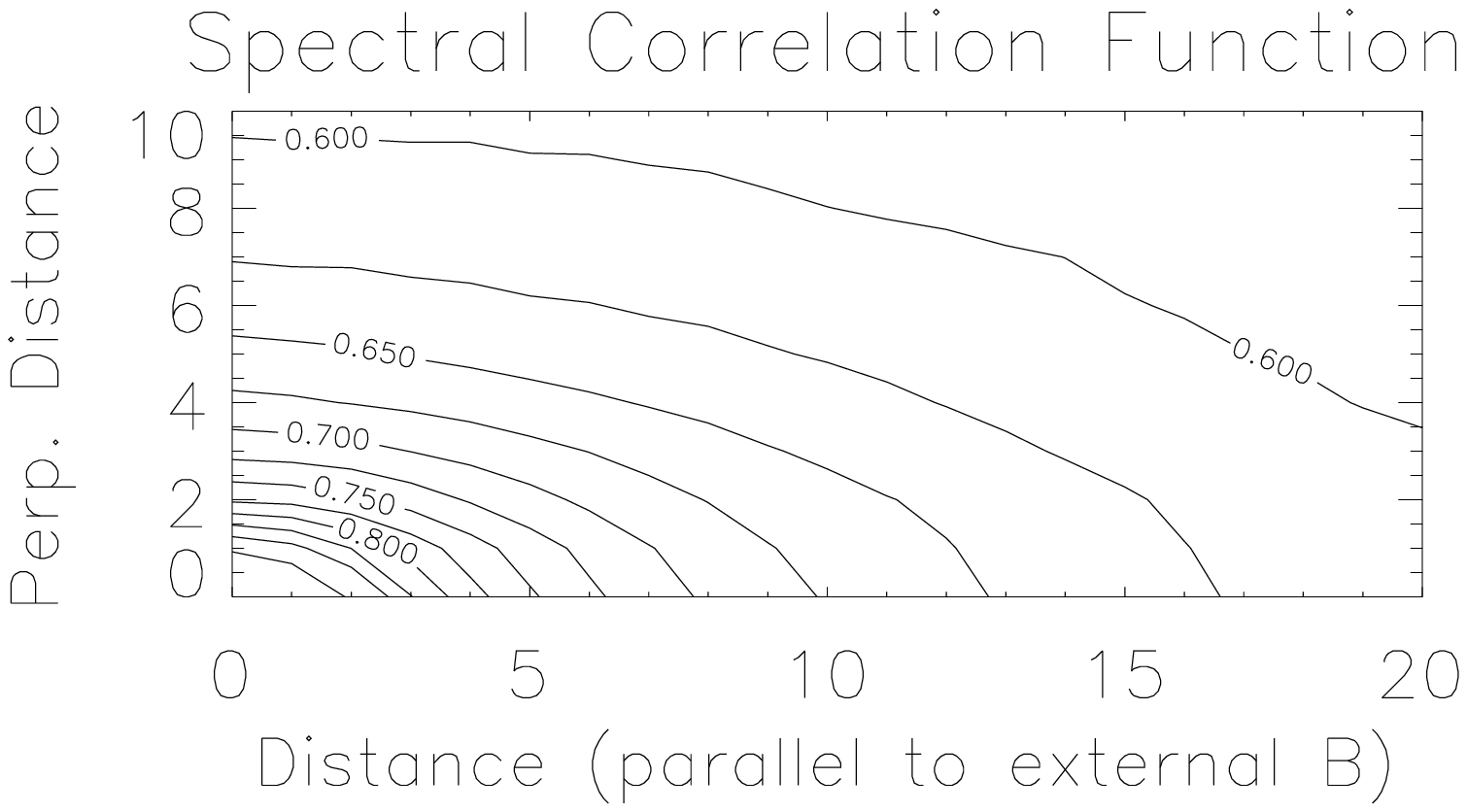} 
}  
\caption{{\it Anisotropies of the synthetic data by Cho, Esquivel} \&
{\it Lazarian}. Anisotropies measured by correlations of
Centroids of Velocity ({\it left})
and by Spectral Correlation Function (SCF)({\it right}) reveal the direction of
magnetic field. Combined with the anisotropy analysis, the SCF (introduced by 
Alyssa Goodman) is likely to become even more useful tool.}  
\end{figure}

Since the anisotropy is related
to the magnetic field, the studies of anisotropy can provide means to analyze
magnetic fields. It is important to study different data sets and channel
maps for the anisotropy. Optical and infrared polarimetry
can benchmark the anisotropies in correlation
functions. We hope that the
anisotropies will reveal magnetic field within dark clouds where grain
alignment and therefore polarimetry fails (see Lazarian 2000,
astro-ph/0003314 for a review of grain alignment).

\section{Summary}

For the first time ever we have an adequate theoretical description of
spectral line statistics in the presence of both velocity and density
fluctuations. This must be exploited to get adequate description of
interstellar turbulence. HI data cubes present a great opportunity for
such a study. To get a better understading of interstellar dynamics we
propose using complementary tools like {\it genus} and {\it anisotropy
analysis} and correlating the statistics of HI with that of molecular
species and ions.

{\bf Acknowledgement}. AL acknowledges the support by the grant
NSF AST-0125544.

\end{document}